# Manifest Verification of QCD Gauge Theory

Yu Kun Qian

*School of Physics, Peking University, Beijing 100871, P.R.China*

We analyze the magnetic moment of gluon, find if QCD is nongauge SU(3) theory then the magnetic moment of gluon varnishes, but if QCD is gauge theory then the magnetic moment of gluon will not vanishes. The magnetic moment of gluon can be measured by investigate the E-M decay of gluball.

QCD gauge theory has run for more than 30 years. Many calculations have been made on QCD dynamics, the results all seems good. But since the complexity of QCD dynamics the only one can be thought is only based on the QCD first principle is the asymptotic freedom theory. So I think it maybe necessary to have a manifest proof to the QCD gauge theory.

I try to investigate the magnetic moment of gluons, try to find the difference between gauge theory and nongauge theory. Fist let look at the magnetic moment of W-meson. We only need the three point vertex $\mathscr{L}_{ww\gamma}$, by simple reasoning we can write   [1]

$$\mathscr{L}_{ww\gamma} = -ie\{(W^+_{\mu\nu}W^\mu A^\nu - W^+_\mu A_\nu W^{\mu\nu}) + \kappa W^+_\mu W_\nu F^{\mu\nu} + \frac{\lambda}{m_w^2} W^+_{\rho\mu} W^\mu{}_\nu F^{\nu\rho} + \tilde{\kappa} W^+_\mu W_\nu \tilde{F}^{\mu\nu} + \frac{\tilde{\lambda}}{m_w^2} W^+_{\rho\mu} W^\mu{}_\nu \tilde{F}^{\nu\rho}\} \qquad (1)$$

$\tilde{\kappa}, \tilde{\lambda}$ terms are CP violating so no contribution to the W-meson magnetic moment. By simple calculation we find

$$\mu_w = \frac{e}{2m_w}(1 + \kappa + \lambda)$$

Now we try to find a three point vertex $\mathscr{L}_{gg\gamma}$ for gluons, neglect the CP violating terms we can write

$$\mathscr{L}'_{gg\gamma} = Tr\{\kappa G_\mu G_\nu F^{\mu\nu} + \frac{\lambda}{\Lambda^2} G_{\rho\mu} G^\mu{}_\nu F^{\nu\rho}\} \qquad (2)$$

If QCD is only SU(3) symmetry not a gauge theory then two terms all could not vanish, if QCD is a gauge theory then the first term has been vanishing. In both cases no conclusion can be made. To distinguish two cases we have to turn to a careful perturbation analysis.

At first we have to mention that Furry theorem [2] is not held for gluon no matter it is gauge theory or not. Since in the gluon case in each vertex there is a $\lambda$ matrix, so for a loop there is a extra factor $Tr(\lambda_1 \lambda_2 ... \lambda_n)$ generally speaking it possibly not equal to $Tr(\lambda_n ... \lambda_2 \lambda_1)$. There is a relation

$$\lambda_i^T = \pm \lambda_i$$

So    Tr($\lambda_n...\lambda_2\lambda_1$)=Tr($\lambda_1^T\lambda_2^T...\lambda_n^T$)=± Tr($\lambda_1\lambda_2...\lambda_n$)

If there is the minus sign then the two diagram with inverse direction of fermion line and odd outer gluon lines will not cancel with each other.

Now we inspect the gg$\gamma$ vertex, suppose QCD is not a gauge theory just SU(3) symmetry. Look at a diagram with two same gluons lines enter and a photon line out, inside the diagram there are several fermion loops and several inner gluon lines. There always has another diagram all the same only all the fermion loops with inversed direction. At first we put the $\lambda$ matrices aside, then just like in the Furry theorem case the two diagrams will cancel with each other. Now we think of the $\lambda$'s, since the entry lines should have the same $\lambda_i$, and for the inner gluon line they always have the same $\lambda_i$ on both ends, then altogether the extra factor is always +1. So the two diagrams will always cancel. This show for nongauge QCD the magnetic moment of gluon vanishes. It's easy to see if we add any number inner lines of $W_\mu, Z_\mu^0 or \gamma$ into the diagram it will not spoil the proof above. After that we look at the QCD gauge theory. Since in QCD gauge theory there are three point and four point vertices the proof above will not hold any more. So in nongauge QCD theory the magnetic moment of gluon vanishes, in QCD gauge theory it not vanish. It will serve as a test for QCD gauge theory. For experiment we can measure the magnetic moment of gluball to see whether it has a magnetic moment.

It is also interesting to consider the contribution of seaquarks inside the gluball, usually they should exist. But seaquarks exist in pairs, if they couple into spin 0 then no magnetic moment of course, if they couple into spin 1 then the spins of quark and antiquark are parallel their magnetic moments will cancel. The magnetic moments produced by trajectory movement should also cancel with each other. So the magnetic moment of gluball only comes from the gluons.

Now we come to the last question: how to measure the magnetic moment of gluball? The answer is very simple, if gluball have no magnetic moment it will have no E-M decay, if it has a E-M decay mode then it's a signal that gluball has magnetic moment. If gluball has a E-M decay mode for example it decay into $e^+e^-$ or $\mu^+\mu^-$ then it will serve as a manifest verification of QCD gauge theory.